# A Bayesian Framework for Community Detection Integrating Content and Link

**Tianbao Yang, Rong Jin**
Dept. of Computer Science and Engineering
Michigan State University
East Lansing, MI 48824

**Yun Chi, Shenghuo Zhu**
NEC Laboratories America
10080 North Wolfe Road
Cupertino, CA 95014

## Abstract

This paper addresses the problem of community detection in networked data that combines link and content analysis. Most existing work combines link and content information by a generative model. There are two major shortcomings with the existing approaches. First, they assume that the probability of creating a link between two nodes is determined only by the community memberships of the nodes; however other factors(e.g. popularity) could also affect the link pattern. Second, they use generative models to model the content of individual nodes, whereas these generative models are vulnerable to the content attributes that are irrelevant to communities. We propose a Bayesian framework for combining link and content information for community detection that explicitly addresses these shortcomings. A new link model is presented that introduces a random variable to capture the node popularity when deciding the link between two nodes; a discriminative model is used to determine the community membership of a node by its content. An approximate inference algorithm is presented for efficient Bayesian inference. Our empirical study shows that the proposed framework outperforms several state-of-the-art approaches in combining link and content information for community detection.

## 1 Introduction

Community detection is an important task in analyzing networked data and it has found applications in a number of domains, such as citation network, online blog network, and the World Wide Web. In this paper, we focus on the problem of combining link and content information for effective community detection. Here, the link information refers to the connections/relationships between nodes in a network. The content information refers to the attributes that describe the properties of individual nodes, where such content information varies from networks to networks. For example, in a citation network, the content of each article is represented by a vector of word histograms; in a co-authorship network, the content is the demographic or affiliation information of individual authors. Our goal is to combine link and content information to identify salient communities in a network.

Several approaches are proposed recently for community detection that combine link and content information. Most of these approaches integrate link information with content information by a generative model. First, they assume that the probability of creating a link between two nodes is determined only by the community memberships of the nodes. It is well known that other factors such as popularity of nodes could also significantly affect the probability of creating links between nodes. Second, in most existing approaches, generative probabilistic models are used to model the content of individual nodes. However, some of the attributes in content may be irrelevant to the community membership assignments. For instance, in citation network, some words in the content of an article are irrelevant to the topic of the article. Therefore these approaches that model contents by a generative approach could be vulnerable to irrelevant attributes.

We propose a Bayesian framework to effectively integrate link information and content information for community detection that explicitly addresses these two shortcomings:

- A new link model is presented that considers *the popularity of nodes* while modeling the link information. As a result, the probability of creating a link between two nodes is jointly decided by their community memberships as well as their popularities.

- A discriminative model is introduced to model the community memberships of nodes by their contents. Therefore irrelevant attributes will be filtered out by the discriminative model.

The rest of the paper is organized as follows. Section 2 reviews the related work for combining link and content analysis for community detection. The proposed Bayesian framework is presented in Section 3 and an efficient algorithm for Bayesian inference is presented Section 4. Section 5 presents the results of our empirical studies on two paper citation networks. Section 6 concludes this work.

## 2   Related Work

Most probabilistic models for combining link and content information are based on topic models. To establish the correspondence of terminology between these models and our model, we note that a topic in topic model corresponds to a community in the paper, topic mixtures correspond to community memberships, and documents correspond to nodes.

In one of the first efforts to combine link information and content information, Cohn and Hoffman [4] combined the PLSA model [7] with PHITS model [3]. It models the link information by a PHITS model, and the content information by a PLSA model. These two models are integrated together by the variables of community memberships that are shared by both models. One shortcoming of this model is that it did not capture the topic relation between citing documents and cited documents. We refer to this combined model as PHITS-PLSA model.

Later on Erosheva et al. [5] extended the LDA model to combine link information and content infromation. It essentially provides a full Bayesian treatment for the PHITS-PLSA model by introducing a Dirichlet prior for topic mixtures. We refer to this model as LDA-Link-Content model.

Recently, Nallapati et al. [11] proposed two combined models. The first model, referred to as Pairwise Link-LDA, combines the mixed membership stochastic block model [1] with the LDA model. To generate a link from document $i$ to document $j$, it first samples a topic variable for each of the two documents, and then generates the link by a Bernolli distribution whose success rate depends on the memberships of both documents. One major limitation of this model is that it has to model the presence and absence of links by a Bernoulli distribution. It is well known that many factors, other than community membership, could result in the absence of a link. The second model in their work, named as Link-PLSA-LDA model, modifies the LDA-Link-Content model. It assumes that the link structure is a bipartite graph with all links emerging from the set of citing documents and pointing to the set of cited documents. It adopts the same generative process for the citing documents as LDA-Link-Content and a PLSA-like model for the cited documents.

Another related work is the latent topic model for hypertext by Gruber et al. [6]. It is built upon the LDA model. In this model, the generation of a link depends on the words in the citing document. When generating a link from a word to documents, instead of generating the target document from a topic-specific distribution over documents, it first samples the target document from a multinomial distribution, and then samples a topic for the the target document. A link is created only when the target document and the word share the same topic.

Besides probabilistic models, several non-probabilistic methods are proposed for combing link and content. Zhu et al. [16] used a matrix-factorization method to induce a new representation of documents from the combination of content and link. The combination of link and content analysis is also studied in the framework of data fusion [15].

## 3   A Bayesian Framework for Community Detection

We denote by $\mathcal{V} = \{1, \cdots, n\}$ the nodes in a network, and by $\mathcal{L}_i = \{j_1, \cdots, j_{N_i}\}$ the set of links starting from node $i$ where $j_\ell$ is the ending node for the $\ell$th link from node $i$. $\mathcal{L} = \bigcup_i \mathcal{L}_i$ includes all the links in the network, and $N = |\mathcal{L}| = \sum_{i=1}^n N_i$ represents the total number of links. We denote by $x_i \in \mathbb{R}^d$ the content vector for node $i$, where $d$ is the number of attributes. $X = (x_1, \ldots, x_n)$ is matrix of size $d \times n$ that includes the content of all nodes. Unlike most existing work on combining content and link information that chooses to model $\Pr(\mathcal{L}, X)$, we propose to model $\Pr(\mathcal{L}|X)$ with no generative process on the contents $X$. The key advantage of this choice is that the proposed model can fit the content information by a discriminative model, thus alleviating the impact of irrelevant attributes. In order to model $\Pr(\mathcal{L}|X)$, we first sample the community variables $z$ for individual nodes by a discriminative model based on the contents $X$; the sampled community variables $z$ will then be combined with the popularity of nodes, denoted by $t$, to create the links between nodes. Below we will describe these two procedures.

## 3.1 A Discriminative Model for Content Analysis

We model the community memberships of nodes based on their contents by a discriminative model. For each node $i$, we first compute its community activation functions $y_k(x_i) = w_k^T x_i, k = 1, \ldots, K$, where $w_k$ is the weight vector for community $k$, and $K$ is the number of communities. We then compute the probability of assigning node $i$ to community $k$ as $\gamma_{ik} = \frac{\exp(y_{ik})}{\sum_l \exp(y_{il})}$. The community variable $z$ that relates to node $i$ will follow the multinomial distribution $Mult(\gamma_{i1}, \ldots, \gamma_{iK})$. Such a model that uses logistic multinomial has been widely used (e.g., [10] and [2]). We can also specify a Gaussian prior for each $w_k$ by $w_k \sim \mathcal{N}(0, \lambda_k^{-1} I)$. Note that by assigning small weights to irrelevant attributes of content and large weights to the informative ones, the discriminative model presented above will be resilient to the noises(i.e., irrelevant attributes) in contents.

We generalize the above linear discriminative model into a nonlinear model by exploring the Gaussian Processes. In particular, we assume the community activation function $y_k(x)$ follows a Gaussian Process, i.e., for any set of points $x_1, \cdots, x_n$, the evaluation of $y(x)$ on these points jointly follows a Gaussian distribution. We denote by the vector $\mathbf{y}_k = (y_{1k}, \ldots, y_{nk})$ the evaluation of $y_k(x)$ at $x_1, \ldots, x_n$. It follows a Gaussian distribution $\mathcal{N}(0, C_k)$, where $C_k \in \mathbb{R}^{n \times n}$ is the covariance matrix specified for community $k$, whose elements $C_k(i, j) = C_k(x_i, x_j) : \mathbb{R}^d \times \mathbb{R}^d \to \mathbb{R}$ is some function on the content of node $i$ and node $j$.

## 3.2 Popularity Driven Link Model

We consider the problem of modeling link $\ell^i$ from node $i$ to node $j$. Given the community membership $\gamma_{ik}$ computed by the discriminative model, we first sample a community variable for node $i$ associated with link $\ell^i$, denoted by $z_\ell^i$. $z_\ell^i \in \{1, \cdots, K\}$ follows the multinomial distribution by $z_\ell^i \sim Mult(\gamma_{i1}, \cdots, \gamma_{iK})$. Unlike the existing work that assumes links are determined only by the community memberships of nodes, we introduce random variable $t_i, i = 1, \ldots, n$ to represent the popularity of each node. A node with high popularity is on average more likely to be linked by the other nodes than a node with a lower popularity. Our assumption is that the probability of creating a link $\ell^i$ from node $i$ to node $j$ is determined by two factors: (a) whether or not both nodes belong to the same community, and (b) the popularity of the ending node $j$. Based on this assumption, we model $\Pr(\ell^i = j | z_\ell^i)$, the probability of creating a link from node $i$ to node

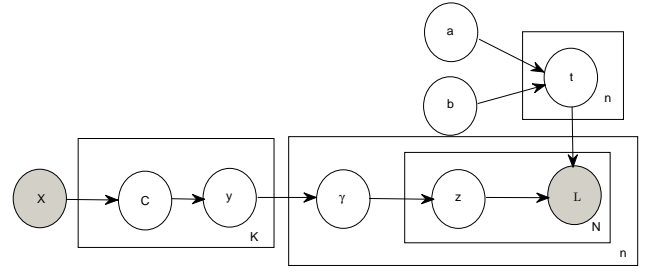

Figure 1: Graphical representation of the generative process for the proposed framework.

$j$ given the community variable $z_\ell^i$, as follows

$$\Pr(\ell^i = j | z_\ell^i) = \frac{t_j \gamma_{jz_\ell^i}}{\sum_{j'} t_{j'} \gamma_{j'z_\ell^i}}$$

$$= \frac{t_j \exp(y_{jz_\ell^i}) / \sum_k \exp(y_{jk})}{\sum_{j'} t_{j'} \exp(y_{j'z_\ell^i}) / \sum_k \exp(y_{j'k})} \qquad (1)$$

Due to the computational reason, we simplify $\Pr(\ell^i = j | z_\ell^i)$ as follows

$$\Pr(\ell^i = j | z_\ell^i) = \frac{t_j \exp(y_{jz_\ell^i})}{\sum_{j'} t_{j'} \exp(y_{j'z_\ell^i})} \qquad (2)$$

We see that the link model in equation (2) approximates equation (1) by assuming that $\sum_l \exp(y_{jl})$ is the same for all node $j$. Finally, we assume that popularity $t_i$ is sampled from a Gamma distribution $Gam(a, b)$ where $a$ and $b$ are the hyper-parameters.

## 3.3 Unified Model for Link and Content Analysis

Below we summarize the generative process for the proposed model that integrates link and content information for community detection. For the discriminative model on the content, we rely on the non-linear model with Gaussian Processes.

1. For each node $i = 1, \cdots, n$, draw $t_i \sim Gam(a, b)$.
2. For each community $k$, draw $\mathbf{y}_k \sim \mathcal{N}(0, C_k)$
3. Compute community memberships $\gamma_{ik} = \exp(y_{ik}) / [\sum_l \exp(y_{il})]$
4. For each node $i = 1, \cdots, n$
   (a) for each link $\ell = 1, \cdots, N_i$, draw $z_\ell^i \sim Mult(\gamma_{i1}, \cdots, \gamma_{iK})$
   (b) draw the target node $j$ from $Mult(\lambda_{1,\ell}^i, \cdots, \lambda_{n,\ell}^i)$ where $\lambda_{j,\ell}^i = t_j \exp(y_{jz_\ell^i}) / \sum_{j'} t_{j'} \exp(y_{j'z_\ell^i})$

Figure 1 shows the graphical representation for the proposed unified framework.

## 4 An Approximate Algorithm for Bayesian Inference

In this section, we present an approximate algorithm for the Bayesian inference of the proposed model, with the focus on using the Gaussian Process as the prior for $\mathbf{y}_k$. We introduce the notations $T = \{t_1, \cdots, t_n\}$, $Z = \{z_1^i, \cdots, z_{N_i}^i, i = 1, \cdots, n\}$, $Y = \{\mathbf{y}_k, k = 1, \cdots, K\}$ to represent the hidden variables for all the nodes in the network. The key to Bayesian inference in our model is to derive the posterior distribution $\Pr(Y|\mathcal{L}, X)$. It requires computing the marginal distribution $\Pr(\mathcal{L}|X) = \sum_Z \int dT dY \Pr(\mathcal{L}, T, Z, Y|X)$, which is in general intractable. We thus approximate the Bayesian Inference by a variational inference. In the following, we omit the conditional variables $X$ in probability notations.

First, we have the joint distribution $\Pr(\mathcal{L}, T, Z, Y)$ expressed as follows:
$$\Pr(\mathcal{L}, T, Z, Y)$$
$$= \prod_{i=1}^{n} \prod_{\ell=1}^{N_i} \Pr(\ell^i = j_\ell | z_\ell^i, T, Y) \Pr(z_\ell^i | Y) \prod_i \Pr(t_i) \prod_k \Pr(\mathbf{y}_k)$$
$$= \prod_{i=1}^{n} \prod_{j \in \mathcal{L}_i} \Pr(\ell^i = j | z_j^i, T, Y) \Pr(z_j^i | Y) \prod_i \Pr(t_i) \prod_k \Pr(\mathbf{y}_k)$$
$$= \prod_{i=1}^{n} \prod_{j \in \mathcal{L}_i} \prod_k \left( \frac{t_j \exp(y_{jk})}{\sum_{j'} t_{j'} \exp(y_{j'k})} \frac{\exp(y_{ik})}{\sum_l \exp(y_{il})} \right)^{z_{jk}^i}$$
$$\times \prod_i Gam(t_i | a, b) \prod_k \mathcal{N}(\mathbf{y}_k | 0, C_k)$$

In the above, we slightly abuse the notation by replacing $z_\ell^i$ with $z_j^i$. Both random variables refer to the community to which node $i$ is assigned for the link $\ell^i$. We also introduce the random variable $z_{jk}^i = I(z_j^i = k)$ where $I(x)$ outputs one when the boolean variable $x$ is true and zero otherwise.

The following lemma provides a lower bound of $\log \Pr(\mathcal{L}, T, Z, Y)$.

**Lemma 1.** The $\log \Pr(\mathcal{L}, T, Z, Y)$ is lower bounded as follows:
$$\log \Pr(\mathcal{L}, T, Z, Y) \geq \log p(\mathcal{L}, T, Z, Y, \eta, \tau)$$
$$= \sum_{i, j \in \mathcal{L}_i} \sum_k z_{jk}^i \left( \log t_j + y_{jk} + y_{ik} \right.$$
$$+ 1 - \frac{1}{\eta_k} \sum_{j'} t_{j'} \exp(y_{j'k}) - \log \eta_k$$
$$+ 1 - \frac{1}{\tau_i} \sum_l \exp(y_{il}) - \log \tau_i \right)$$
$$+ \sum_j \log Gam(t_j | a, b) + \sum_k \log \mathcal{N}(\mathbf{y}_k | 0, C_k)$$

where $\eta = (\eta_1, \cdots, \eta_K)$, $\tau = (\tau_1, \cdots, \tau_n)$ are the variational parameters.

The above lemma follows directly from $-\log x \geq 1 - x$.

In variational inference, we approximate the posterior of $\Pr(T, Z, Y|\mathcal{L})$ by a factorized distribution of $q(T)q(Z)q(Y)$, and bound the marginal probability $\Pr(\mathcal{L})$ as
$$\log \Pr(\mathcal{L}) \geq \sum_Z \int dT dY q(T)q(Z)q(Y) \log \frac{p(\mathcal{L}, T, Z, Y, \eta, \tau)}{q(T)q(Z)q(Y)} \tag{3}$$
The optimal solution of $q(T)$, $q(Z)$ and $q(Y)$ is obtained by maximizing the above lower bound by alternating optimization. In the following, we present the solution to each of the three approximated posterior distributions given the other two distributions are fixed.

With fixed $q(Z)$ and $q(Y)$, the optimal solution to $q(T)$ is computed by
$$\log q(T) = E_{Z,Y} \log p(\mathcal{L}, T, Z, Y, \eta, \tau) + c$$
where $c$ is an appropriate normalization constant. With the details omitted, the explicit form of $q(T)$ is given by
$$q(T) = \prod_j q(t_j) = \prod_j Gam(t_j | \tilde{a}_j, \tilde{b}_j)$$
where
$$\tilde{a}_j = a + \sum_{i \in \mathcal{I}(j)} \sum_k E[z_{jk}^i]$$
$$\tilde{b}_j = b + \sum_k \sum_{(i', j' \in \mathcal{E}_i)} E[z_{j'k}^{i'}] \frac{E[\exp(y_{jk})]}{\eta_k}$$
where $\mathcal{I}(j)$ denotes the set of nodes that are linked to node $j$.

With fixed $q(T)$ and $q(Y)$, the optimal solution to $q(Z)$ is computed by
$$\log q(Z) = E_{T,Y} \log p(\mathcal{L}, T, Z, Y, \eta, \tau) + c$$
and the explicit form of $q(Z)$ is given by
$$q(Z) = \prod_i \prod_{j \in \mathcal{E}_i} Mult(z_j^i | \psi_{j1}^i, \cdots, \psi_{jK}^i)$$
where
$$\psi_{jk}^i \propto$$
$$\exp \left( E[y_{jk}] + E[y_{ik}] - \frac{\sum_{j'} E[t_{j'}] E[\exp(y_{j'k})]}{\eta_k} - \log \eta_k \right)$$

With fixed $q(T)$ and $q(Z)$, the solution to $q(Y)$ is computed by

$$\log q(Y) = E_{T,Z} \log q(\mathcal{L}, T, Z, Y, \eta, \tau) + c$$

However, due to the exponential terms in $\log q(\mathcal{E}, T, Z, Y)$ about Y, we cannot have a closed solution for $q(Y)$. Instead, we assume a Gaussian factorization of $q(Y)$ by $q(Y) = \prod_k \mathcal{N}(\mathbf{y}_k | \mathbf{m}_k, \Sigma_k)$ and maximize the lower bound in (3) with regard to $\mathbf{m}_k$ and $\Sigma_k$, i.e.,

$$\max_{q(Y)} E_{T,Z,Y} \quad \log p(\mathcal{L}, T, Z, Y, \eta, \tau) - \int dY q(Y) \log q(Y) \tag{4}$$

The algorithm for this maximization problem is given in the Appendix.

Finally, we note that the solution of $q(Z)$, $q(T)$, and $q(Y)$ depend on the variational parameters $\eta, \tau$. We can take a variational EM algorithm to solve this problem. In the variational E-step, we compute $q(T)$, $q(Z)$ and $q(Y)$ given $\eta, \tau$; in the variational M-step, we estimate $\eta$ and $\tau$ by maximizing the lower bound in (3), i.e.,

$$\max_{\eta, \tau} E_{T,Z,Y} \log p(\mathcal{E}, T, Z, Y, \eta, \tau)$$

The optimal solution of $\eta, \tau$ can be shown as

$$\eta_k = \sum_{j'} E[t_{j'}] E[\exp(y_{j'k})]$$

$$\tau_i = \sum_l E[\exp(y_{il})]$$

Once we obtain the posterior distribution $q(Y) = \prod_k \mathcal{N}(\mathbf{y}_k | \mathbf{m}_k, \Sigma_k)$, the marginal distribution of each $y_{ik}$ is also a Gaussian distribution with $q(y_{ik}) = \mathcal{N}(y_{ik} | m_{ik}, \Sigma_{k,ii})$, where $m_{ik}$ is the $i$th element in $m_k$, and $\Sigma_{k,ii}$ is the $i$th diagonal element of $\Sigma_k$. $\gamma_{ik}$, the probability of assigning a node $i$ to community $k$ is computed as

$$\gamma_{ik} = \int \frac{\exp(y_{ik})}{\sum_l \exp(y_{il})} \mathcal{N}(y_{ik} | m_{ik}, \Sigma_{k,ii}) dy_{i1} \cdots dy_{iK}$$

which can be computed by a sampling method or approximated by the Extended MacKay approach[8] given by:

$$\gamma_{ik} \sim \frac{\exp(\kappa(\Sigma_{k,ii}) m_{ik})}{\sum_l \exp(\kappa(\Sigma_{l,ii}) m_{il})}$$

where $\kappa(x) = 1/\sqrt{1 + \pi x/8}$.

# 5 Experiments

In this section, we present the empirical study results by comparing our model with several state-of-the-art approaches. We first describe the data sets, the metrics, and the baselines used for the evaluation.

## 5.1 Data Sets

In the experiments, we used two paper citation networks, namely Cora Data Set, and Citeseer Data Set.

**Cora Data Set** is a subset of the larger Cora citation data set [9] collected by Lise Getoor's research group. This data set has been widely used for classification, clustering, and studies of combining link and content. It contains 2708 scientific publications that are classified into seven machine learning areas. These articles are linked by a total of 5429 citations to form the citation network. Each paper corresponds to a node and is described by a 0/1-valued word vector indicating the absence/presence of the corresponding word. The number of unique words is 1433.

**Citeseer Data Set** is a subset of the larger Citeseer data set[1] also collected by Lise Getoor's research group. The Citeseer data set consists of 3312 scientific publications and 4732 links. Each publication classified into one of the six topics is described by a 0/1 valued word vector. The dictionary consists of 3703 unique words.

For the number of communities, i.e. K, without further evidence, we set K to be the number of class labels in the ground truth. For Cora data set, K is set to 7 because papers are labeled as one of the seven classes; for Citeseer data set, K is set to 6 since papers are classified as one of the six topics.

## 5.2 Metrics

We use three metrics for evaluation where these metrics are commonly used in clustering and community detection. These metrics are *normalized mutual information(NMI), pairwise F-measure(PWF), modularity(Modu)*. *NMI* and *PWF* are computed with the ground truth of community labels, and *Modu* is computed without the ground truth of the community labels(i.e., class or topic). Modularity is a commonly used metric in social network analysis to evaluate methods for community detection. It measures the goodness of a community structure in terms of links, which can be used as an evidence for the performance of our link model. We emphasize that although modularity is not an ideal evaluation metric, it makes it possible for us to compare our work to the existing ones since it is such a widely used metric.

With the ground truth of community labels for each node, we can form the true community structure $V = \{V_1, \ldots, V_K\}$, where $V_k$ contains the set of nodes that are in the $k$th community. Assume the community structure given by the algorithms is represented by $V' = \{V'_1, \ldots, V'_K\}$, then the *mutual information* between the two is defined as

$$\widehat{MI}(V, V') = \sum_{V_i, V'_j} p(V_i, V'_j) \log \frac{p(V_i, V'_j)}{p(V_i)p(V'_j)}$$

and the *normalized mutual information* is defined by

$$NMI(V, V') = \frac{\widehat{MI}(V, V')}{\max(H(V), H(V'))}$$

where $H(V)$ and $H(V')$ are the entropies of the partitions $V$ and $V'$. The higher the normalized mutual information, the closer the partition is to the ground truth.

To compute the *pairwise F-measure*, let $T$ denote the set of node pairs that have the same label, $S$ denote the set of node pairs that are assigned to the same community, $|T|$ denote the cardinality of set $T$. The *pairwise F-measure* is computed from the pairwise precision and recall, as the following

$$precision = |S \bigcap T|/|S| \quad recall = |S \bigcap T|/|T|$$

$$PWF = \frac{2 \times precision \times recall}{precision + recall}$$

The higher the $PWF$, the better is the partition.

*Modularity* is proposed by Newman et al. [13] for measuring community partitions. For a given community partition $V = \{V_1, \ldots, V_K\}$, the modularity is defined as

$$Modu(V) = \sum_k \left[ \frac{Cut(V_k, V_k)}{Cut(V, V)} - \left( \frac{Cut(V_k, V)}{Cut(V, V)} \right)^2 \right]$$

where $Cut(V_i, V_j) = \sum_{p \in V_i, q \in V_j} w_{pq}$. As stated in [13], modularity measures how likely a network is generated due to the proposed community structure versus generated by a random process. Therefore, a higher modularity value indicates a community structure that better explains the observed network.

### 5.3 Baselines

The following baselines are used in our evaluation:

**PHITS-PLSA** is the model proposed by Cohn et al. [4]. The community membership of each node for this model is given by the topic mixture of each paper document. We run this algorithm until the difference of log-likelihood between consecutive steps is within $10^{-8}$. The combination coefficient of $\alpha$ is tuned to obtain the optimal $NMI$ and $PWF$.

**LDA-Link-Content** is the mixed membership model proposed by Erosheva et al. [5]. The community membership of each node is given by the mean of posterior for the topic mixture of corresponding document. We run the variational EM algorithm for this model until the difference of the lower bound of the log-likelihood is within $10^{-8}$.

**Link-Content-Factorization(LCF)** is the approach based on matrix factorization proposed by Zhu et al. [16]. In the method, a new low-dimensional representation of each document is computed from the content and link information. K-means algorithm is then applied to the derived representation to get the communities. The parameters in the method are tuned to obtain the optimal $NMI$ and $PWF$.

**Spectral Clustering with fused kernel** This baseline is motivated by the data fusion method for clustering [15]. We compute a kernel from the content and then we combine this kernel matrix with the adjacent matrix derived from the link structure in a linear manner. This combined matrix is then used as a new similarity matrix for spectral clustering. The spectral clustering method used is the Normalized Cut algorithm presented in [14]. A RBF kernel is used to construct the similarity matrix for content. The parameters in RBF kernel, and the combination coefficient are tuned to obtain the optimal $NMI$ and $PWF$. We refer to the algorithm as NCUT.

Finally, for the purpose of presentation, we refer our model as the **C**ombined **P**opularity-driven **L**ink model and **D**iscriminative **C**ontent model(**C-PLDC**). The hyperparameters $a, b$ for the Gamma prior of $t$ are set to very lower values to enforce an uninformative prior ($a = b = 10^{-3}$). The covariance matrix we used for all communities are the same. We present results for RBF covariance matrix, which is given as $C_k(i, j) = \theta \exp(-\|x_i - x_j\|^2/2\sigma^2)$. The parameters for the covariance matrix are tuned to maximize the lower bound of the log-likelihood. To prevent numerical problems, we add a noise level or "jitter" term $\nu \delta_{i,j}$ to the covariance function as recommended by [12], where $\nu$ is set to $10^{-5}$. The variational EM algorithm is run until the difference of the lower bound of the log-likelihood is within $10^{-8}$. The final community memberships are computed by Extended MacKay approach.

The results for Cora and Citeseer data sets are shown in Tables 1 and 2, respectively. We can see that our model performs better than other baselines. In order to verify if the introduction of popularity is necessary for modeling links, we include the results of our model by setting the popularity of all nodes to a constant 1, denoted by C-PLDC(t=1). It is clear that the pro-posed framework with popularity driven link model performs noticeably better than the one without popularity. In addition, comparing to the four baseline methods, we observe that C-PLDC(t=1), the proposed model without modeling popularity, performs significantly better than the baseline models. Since the key component of C-PLDC(t=1) is to model content information by a discriminative analysis, this comparison further confirms that the introduction of discriminative model for content analysis is beneficial for community detection. Finally, we show in Figure 2(a) and 2(b) the change in NMI(similar results for PWF and Modu) when we vary the parameters $\sigma^2$ and $\theta$ in computing the covariance matrix. We also include in the two figures the change in lower bound of the log-likelihood function computed by the proposed approximate algorithm. It is interesting to observe that both curves are consistent with each other in overall trends, which makes it possible to determine the optimal parameter for $\sigma^2$ and $\theta$.

Table 1: Evaluation on Cora dataset

| Algorithm | NMI | PWF | Modu |
|---|---|---|---|
| PHITS-PLSA | 0.3140 | 0.3526 | 0.3956 |
| LDA-Link-Content | 0.3587 | 0.3969 | 0.4576 |
| LCF | 0.2421 | 0.2780 | 0.2802 |
| NCUT | 0.2444 | 0.3062 | 0.3703 |
| C-PLDC(t=1) | 0.4294 | 0.4264 | 0.5877 |
| C-PLDC | **0.4887** | **0.4638** | **0.6160** |

Table 2: Evaluation on Citeseer dataset

| Algorithm | NMI | PWF | Modu |
|---|---|---|---|
| PHITS-PLSA | 0.1188 | 0.2596 | 0.3863 |
| LDA-Link-Content | 0.1920 | 0.3045 | 0.5058 |
| LCF | 0.1416 | 0.2621 | 0.3090 |
| NCUT | 0.1592 | 0.2957 | 0.4280 |
| C-PLDC(t=1) | 0.2303 | 0.3340 | 0.5530 |
| C-PLDC | **0.2756** | **0.3611** | **0.5582** |

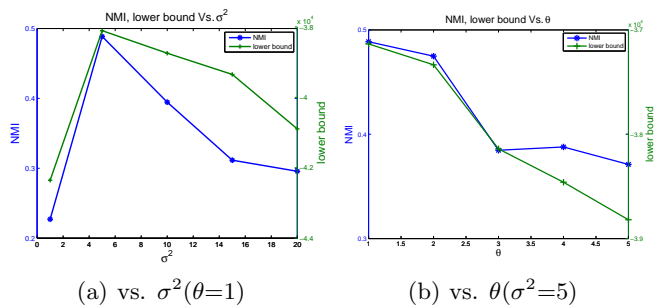

(a) vs. $\sigma^2(\theta=1)$          (b) vs. $\theta(\sigma^2=5)$

Figure 2: NMI and lower bound of log-likelihood vs. $\sigma^2$ and $\theta$. (a) fix $\theta = 1$, change $\sigma^2$; (b) fix $\sigma^2 = 5$, change $\theta$

# 6 Conclusion

In this paper, we present a new Bayesian framework to combine link and content for community detection in networked data. Unlike most probabilistic models of combining link and content by a generative process on link and content, we propose a discriminative model on the content which is extended to the Gaussian Process. To more accurately model the link, we introduce new variables to capture the popularity of nodes. Empirical studies show that our model significantly outperforms other state-of-the-art approaches. For future work, we may consider other types of network data rather than citation networks presented in the paper; we would also compare our model to other models on the link prediction accuracy.

## Appendix: Computation of $q(Y)$

In this appendix, we present an efficient algorithm to solve the problem as in (4). We assume $q(Y)$ can be factorized as $q(Y) = \prod_k \mathcal{N}(\mathbf{y}_k | \mathbf{m}_k, \Sigma_k)$. By noting that $E[y_{ik}] = m_{ik}$, $E[\exp(y_{ik})] = \exp(m_{ik} + \Sigma_{k,ii}/2)$, the problem in equation (4) is reduced to:

$$\max_{m_k, \Sigma_k} \sum_k \mathbf{p}_k^T \mathbf{m}_k - \sum_k \mathbf{q}_k^T \exp(\mathbf{m}_k + \mathbf{s}_k/2)$$

$$+ \sum_k \frac{1}{2} \left[ \log \frac{|\Sigma_k|}{|C_k|} + n - Tr(C_k^{-1}\Sigma_k) - \mathbf{m}_k^T C_k^{-1} \mathbf{m}_k \right]$$

where

$$
\begin{aligned}
p_{ik} &= \sum_{j \in \mathcal{I}(i)} E[z_{ik}^j] + \sum_{j \in \mathcal{E}_i} E[z_{jk}^i] \\
q_{ik} &= \frac{\sum_{(i',j' \in \mathcal{E}_i)} E[z_{j'k}^{i'}] E[t_i]}{\eta_k} + \frac{\sum_{j \in \mathcal{E}_i, l} E[z_{jl}^i]}{\tau_i} \\
s_{ik} &= \Sigma_{k,ii}
\end{aligned}
$$

Since $\mathbf{m}_k$ is coupled with $\Sigma_k$, below we describe a coordinate descent algorithm for updating $\mathbf{m}_k$ and $\Sigma_k$ iteratively.

By fixing $\Sigma_k$, our problem for $\mathbf{m}_k$ becomes

$$\max_{m_k} \sum_k \mathbf{p}_k^T \mathbf{m}_k - \sum_k \hat{\mathbf{q}}_k^T \exp(\mathbf{m}_k) - \frac{1}{2} \mathbf{m}_k^T C_k^{-1} \mathbf{m}_k$$

where $\hat{\mathbf{q}}_k = \mathbf{q}_k \cdot \exp(\mathbf{s}_k/2)$. This is a convex optimization problem, and can be solved efficiently by Newton-Raphson method. In particular, we compute the gradient and Hessian matrix as

$$
\begin{aligned}
\mathbf{g}_k &= \mathbf{p}_k - \hat{\mathbf{q}}_k \cdot \exp(\mathbf{m}_k) - C_k^{-1} \mathbf{m}_k \\
H_k &= -W_k - C_k^{-1}
\end{aligned}
$$

where $W_k$ is diagonal matrix with $W_{k,ii} = \hat{q}_{ik} \exp(m_{ik})$. $\mathbf{m}_k$ is updated as $m_k = m_k + H_k^{-1} \mathbf{g}_k$, i.e.,

$$\mathbf{m}_k = \mathbf{m}_k + (W_k + C_k^{-1})^{-1} (\mathbf{p}_k - \hat{\mathbf{q}}_k \cdot \exp(\mathbf{m}_k) - C_k^{-1} \mathbf{m}_k)$$

The matrix inverse $W_k + C_k^{-1}$ can be computed efficiently by the inversion lemma, i.e.,

$$(W_k + C_k^{-1})^{-1} = C_k - C_k W_k^{1/2} B_k^{-1} W_k^{1/2} C_k$$

where $B_k = I + W_k^{1/2} C_k W_k^{1/2}$ is guaranteed to be well-conditioned. Let $\mathbf{v}_k = W_k \mathbf{m}_k + \mathbf{p}_k - \hat{\mathbf{q}}_k \cdot \exp(\mathbf{m}_k)$, then the update formula can be written as

$$\mathbf{m}_k = C_k(\mathbf{v}_k - W_k^{1/2} B_k^{-1} W_k^{1/2} C \mathbf{v}_k) = C_k \mathbf{u}_k$$

Another benefit is that we can calculate $\mathbf{m}_k C_k^{-1} \mathbf{m}_k = \mathbf{m}_k^T \mathbf{u}_k$ without computing $C_k^{-1}$.

By fixing $\mathbf{m}_k$, the problem for $\Sigma_k$ becomes

$$\max_{\Sigma_k} -\tilde{\mathbf{q}}_k^T \exp(\mathbf{s}_k/2) + \frac{1}{2} \left[ \log \frac{|\Sigma_k|}{|C_k|} - Tr(C_k^{-1}\Sigma_k) \right]$$

where $\tilde{\mathbf{q}}_k = \mathbf{q}_k \cdot \exp(\mathbf{m}_k)$. To simplify our computation, we assume that $\Sigma_k$ share the same eigenvectors as $C_k$. Let $(\sigma_{ki}, \mathbf{v}_{ki}), i = 1, \dots, n$ be the eigenvalues and eigenvectors of matrix $C_k$. We then have $\Sigma_k = \sum_i \lambda_{ki} \mathbf{v}_{ki} \mathbf{v}_{ki}^T$ where $\lambda_{ki}$ is the combination weight that needs to be optimized. The related optimization problem becomes

$$\max_{\lambda_k} -\tilde{\mathbf{q}}_k^T \exp(\tau_k \circ \lambda_k) + \frac{1}{2} e^T \log(\lambda_k) - \frac{1}{2} e^T (\lambda_k./\sigma_k)$$

where $\tau_k = (|\mathbf{v}_{k1}|_2^2, \dots, |\mathbf{v}_{kn}|_2^2)$, $\log(\lambda_k) = (\log(\lambda_{k1}), \dots, \log(\lambda_{kn}))$, and $\lambda_k./\sigma_k = (\lambda_{k1}/\sigma_{k1}, \dots, \lambda_{kn}/\sigma_{kn})$. This problem is again a convex optimization problem and can be solved efficiently by Newton-Raphson method.